\documentclass{aa}
\usepackage{graphicx}
\usepackage{times}
\usepackage{latexsym}
\usepackage{epsfig}
\usepackage{txfonts}
\usepackage{natbib}
\usepackage{supertabular,multirow,longtable,rotating}
\usepackage{amssymb}
\usepackage{color}

\bibpunct{(}{)}{;}{a}{}{,} 

\begin{document}

   \title{Linear/circular spectropolarimetry of diffuse interstellar bands~\thanks{%
   Based on observations obtained at the Canada-France-Hawaii Telescope (CFHT) which is 
   operated by the National Research Council of Canada, the Institut National des Sciences de l'Univers
   of the Centre National de la Recherche Scientique of France, and the University of Hawaii.}}

   \author{N.L.J.~Cox \inst{1}
   	   P.~Ehrenfreund \inst{2}
	   B.H.~Foing  \inst{3} 
	   L.~d'Hendecourt \inst{4}
	   F.~Salama \inst{5}
	   P.J.~Sarre \inst{6}
	      }
	
   \institute{Institute of Astronomy, K.U.Leuven, Celestijnenlaan 200D, 3001 Leuven
             \and
	     Astrobiology Group, Leiden Institute of Chemistry, Leiden University, NL
	     \and
	     ESA Research and Scientific Support Department, P.O. Box 299, 2200-AG~~Noordwijk, The Netherlands
	     \and
	     Institut d'Astrophysique Spatiale, France
	     \and
	     NASA Ames Research Centre, CA, USA
	     \and
	     School of Chemistry, The University of Nottingham, University Park, Nottingham, NG7 2RD, UK
	      }

   \offprints{N.L.J.~Cox, \email{nick@ster.kuleuven.be}}

   \date{Received 20 December 2010; accepted 22 March 2011}

   \abstract{The identification of the carriers of diffuse interstellar bands (DIBs) remains one of the 
   long-standing mysteries in astronomy. 
   The detection of a polarisation signal in a DIB profile can be used to distinguish between a dust 
   or gas-phase carrier. The polarisation profile can give additional information on the grain or 
   molecular properties of the absorber.
   }{%
   To measure the polarisation efficiency of the carriers of the diffuse interstellar bands.
   }{%
   In order to detect and measure the linear and circular polarisation of the DIBs 
   we observed reddened lines of sight showing continuum polarisation.
   For this study we selected two stars \object{HD197770} and \object{HD 194279}.
   We used high-resolution ($R \sim 64\,000$) spectropolarimetry in the wavelength range from 3700 to 10480~\AA\ with 
   the ESPaDOnS \'{e}chelle spectrograph mounted at the CFHT.
   }{%
   High S/N and high resolution Stokes V (circular), Q and U (linear) spectra were obtained.
   We constrained upper limits by a factor of 10 for previously observed DIBs.
   Furthermore, we analysed $\sim$30 additional DIBs for which no spectropolarimetry data has been obtained before.
   This included the 9577~\AA\ DIB and the 8621~\AA\ DIB.
   The former is attributed to the $C_{60}^+$ fullerene, which could become aligned in a magnetic field. 
   The latter shows a tight correlation with the amount of dust in the line-of-sight and therefore most likely may show 
   a polarisation signal related the aligned grains.
   }{%
   The lack of polarisation in 45 DIB profiles suggests that none of 
   the absorption lines is induced by a grain-type carrier. The strict upper limits, less than $\sim$0.01\%, derived for
   the observed lines-of-sight imply that if DIBs are due to gas-phase molecules these carriers have 
   polarisation efficiencies which are at least 6 times, and up to 300 times, smaller than those predicted 
   for grain-related carriers.
     }


   \titlerunning{Linear and circular spectropolarimetry of diffuse interstellar bands}
   \authorrunning{Cox, Ehrenfreund, Foing, d'Hendecourt, Salama \& Sarre}

   \maketitle

\section{Introduction}\label{sec:intro}
The diffuse interstellar bands (DIBs) are more than 300 absorption lines in the optical 
spectrum that reside in the interstellar medium (\citealt{1934PASP...46..206M}, \citealt{1995ARA&A..33...19H}).
See for example \citet{2008ApJ...680.1256H} for a recent DIB inventory.
DIBs are ubiquitously present throughout the Galaxy and they have been detected also in other galaxies
(\citealt{2002ApJ...576L.117E}, \citealt{2005A&A...429..559S}, \citealt{2006A&A...447..991C}, \citealt{2007A&A...470..941C}, 
\citealt{2008A&A...485L...9C}, \citealt{2008A&A...480L..13C}, \citealt{2008A&A...492L...5C}).
Not a single carrier has been identified unambiguously yet. 
Their relative large widths argue against atoms and di-atomic molecules in the gas-phase. 
And although their intensity is related to the extinction by dust grains their (spectral) properties and behaviour 
are more consistent with large gas-phase molecules (see also the review by \citealt{2006JMoSp.238....1S}).
In particular the substructure in several DIB profiles indicates that the carrier(s) are large gas phase molecules 
(\citealt{1995MNRAS.277L..41S}, \citealt{1996A&A...307L..25E}, \citealt{2004ApJ...611L.113C}).
DIBs respond to the local environmental conditions, in particular to the effective strength of the UV field 
(\emph{e.g.}\ \citealt{1997A&A...326..822C}, \citealt{2006A&A...447..991C}).
Their strength variation could reflect the local charge state balance of the carrier molecules
(\citealt{2005A&A...432..515R}, \citealt{Cox2006b}).
Therefore, specific groups of stable UV resistant molecules (such as PAHs, fullerenes and carbon chains) are commonly 
postulated carrier candidates (\citealt{1995ARA&A..33...19H}). 
Interstellar grains are known to become aligned when situated in a magnetic field which is evidenced by linear and circular
continuum polarisation (\emph{e.g.}\ \citealt{1965ApJ...141.1340S}).
The linear continuum polarisation can be described by the following empirical relation:
\begin{equation}
P(\lambda) = P_{\rm max}\ {\rm exp}(-K\ {\rm ln}^2(\lambda_{\rm max} / \lambda)), \label{eq:serkowski} 
\end{equation}
with $K(\lambda_{\rm max})$ = (1.66$\pm$0.09) $\lambda_{\rm max}$ + (0.01$\pm$0.05) (\citealt{1965ApJ...141.1340S}, 
\citealt{1974AJ.....79..581C}, \citealt{1992ApJ...386..562W}).
The wavelength dependency of the polarisation is mainly determined by the composition and size of the dust particles.
The polarisation efficiency, $P(\lambda)/A(V)$, is a function of various factors, such as porosity and shape (Voshchinnikov \& Das 2008).

For example, it has been well established that the silicate features at 9.7 and 18.5~$\mu$m show an excess of polarisation 
(\citealt{1998MNRAS.299..743A}, \citealt{2000MNRAS.312..327S}). 
Also ice-features show polarisation (3.1~|$\mu$m O-H stretching mode of water ice; \citealt{1996ApJ...461..902H},
or ice features near 4.6-4.7~$\mu$m show polarization due to CO and CN-bearing species; \citealt{1996ASPC...97..243C}).
These detections have been taken as evidence for alignment of core/mantle grains in molecular clouds.
Thus, polarization of the 3.4~$\mu$m C-H feature is expected if it is due to carbonaceous mantles on silicate cores.
However, no polarisation has been detected for the 3.4~$\mu$m feature (\citealt{2006ApJ...651..268C}).
\citet{1999ApJ...512..224A} obtained an upper limit of $\sim$0.06$\pm$0.13\% for $\Delta p$, which is a factor 5 below the predicted value
for $\Delta p_{9.7}/\tau_{3.4}$ of 0.4\%.

Several theoretical and experimental studies predict that also large (ionized) molecules, such as PAHs and fullerenes, 
can align, via for example the Barnett effect, under certain physical conditions (see \emph{e.g.}\ \citealt{1990ApJ...356..507B}, 
\citealt{1992A&A...253..498R}, \citealt{1994MNRAS.268..713L}, \citealt{1997ApJ...478..395W}).
Depending on the polarisation (\emph{i.e.}\ parallel versus perpendicular) of the incident light changes can be seen  
in the electronic absorption spectra of large molecules that have some intrinsic asymmetry.
A summary of different proposed alignment mechanisms is given in {\citet{2007A&A...465..899C}}.
Therefore, the polarisation signal across a DIB profile could provide further constraints on the (molecular) properties of their carriers.
Spectropolarimetry of diffuse bands has been limited to about ten lines-of-sight and only nine individual DIBs, although no
DIB polarisation has yet been detected. Recent studies include those by
\citet{1992ApJ...398L..69A,1995ApJ...448L..49A}, \citet{1996ASPC...97..143S}, and \citet{2007A&A...465..899C} 
(but see also references therein for earlier work).
The most recent and comprehensive study involved three sightlines and six DIBs \citep{2007A&A...465..899C}.
This study set the most stringent detection limits, between 0.01 and 0.14\%, for linear and circular polarisation of 6 narrow DIBs.
These values exclude classical grains as carriers of the $\lambda\lambda$~5780, 5797, 6613 and 6284 DIBs.
That the 6379 and 6613~\AA\ DIBs originate from (classical) grains could only be marginally excluded from
previous polarisation measurements.
This lack of line polarisation of DIBs implies that the DIB carriers are not embedded in or attached onto large - silicate - grains 
(\emph{i.e.} those that produce optical extinction and polarisation), but might still be related to smaller - carbonaceous - grains, 
\emph{i.e.} those that produce the far-UV extinction. The current constraint on the line polarisation is still consistent with a gas 
phase carrier for which the polarisation signal could be very weak.

The aim of the present study is to ascertain whether or not the DIB carriers can give rise to an observable polarisation
and what that means for their identity.
There is not {\it a priori} way of knowing if and which DIB carriers are related to grains or molecules, and therefore
each DIB could or could not give rise to significant line polarisation predicted for grains or very weak polarisation from molecules.
Note that only a few DIBs exhibit a strong correlation with each other thus indicating that the majority of the DIBs have different,
though possibly physically/chemically related, carriers (\citealt{1997A&A...326..822C}, \citealt{2010ApJ...708.1628M}).
We present new observations for two lines-of-sight previously studied but with an order of magnitude higher sensitivity
and for many more additional individual DIBs not included in spectropolarimetry studies before.
In particularly, our study includes also weak DIBs and DIBs in the near-IR.
    
\begin{table}[th!]
\caption{Target and line-of-sight information (from literature).}
\centering
\begin{tabular}{lll}\hline\hline
ID					& \object{HD 197770}			& \object{HD 194279}		\\ \hline
Ra (J2000)\tablefootmark{a}		& 20:43:13.68				& 20:23:18.16			\\
Dec (J2000)\tablefootmark{a}    	& +57:06:50.4				& +40:45:32.6			\\
Spectral Type\tablefootmark{a}		& B2\,III				& B1.5\,Ia			\\
distance (pc)				& $\sim$440\tablefootmark{g} 		& 1100\tablefootmark{i}		\\
$B$ (mag)\tablefootmark{a}		& 6.594    				& 7.92     			\\
$V$ (mag)\tablefootmark{a}		& 6.341    				& 7.09    			\\
\ensuremath{{\mathrm E}_{\mathrm (B-V)}}\ (mag)	& 0.58$\pm$0.04\tablefootmark{c}& 1.21 				\\
$A_V$ (mag)				& 1.61$\pm$0.14\tablefootmark{c}	& 3.9				\\
$R_V$					& 2.8\tablefootmark{d}, 2.77$\pm$0.15\tablefootmark{c}	& 3.2\tablefootmark{d}, 3.25$\pm$0.03\tablefootmark{e} 	\\
$P_{\rm max}$ (\%)			& 3.83\tablefootmark{b}, 3.81\tablefootmark{f}		& 2.77		\\ 
$\lambda_{\rm max}$ ($\mu$m)    	& 0.49\tablefootmark{b}, 0.51\tablefootmark{f}		& 0.58		\\
$P_{\rm max}/A_V$ (\%/mag)		& 2.4 					& 0.71				\\ 
$P_{\rm max}/\tau_V$\tablefootmark{h}	& 0.026					& 0.008				\\
\hline				
\label{tb:targets}
\end{tabular}
\tablefoot{
\tablefoottext{a}{From Simbad}
\tablefoottext{b}{From \citet{1995ApJ...445..947C}}
\tablefoottext{c}{From \citet{2004ApJ...616..912V}}
\tablefoottext{d}{Computed from $R_V = 5.5 \lambda_{\rm max}$ (\citealt{1975ApJ...196..261S})}
\tablefoottext{e}{From \citet{2003AN....324..219W}}
\tablefoottext{f}{From \citet{1993ApJ...403..722W}}
\tablefoottext{g}{From \citet{1994ApJS...95..419D}}
\tablefoottext{h}{The optical depth $\tau_V = A_V/1.086$.}
\tablefoottext{i}{Spectroscopic parallax distance (\citealt{2002ApJ...567..391M}); see also Sect.~\ref{sec:hd194279}.}
}
\end{table}

\section{Spectropolarimetric observations}\label{sec:observations}
For the present study we obtained new spectropolarimetry data with ESPaDOnS at the Canadian-French-Hawaian Telescope (CFHT). 
The data were taken on 8-9 July and 21-25 July 2008 under good seeing ($\leq$ 1\arcsec) conditions.
ESPaDOnS is a high-resolution high-efficiency 2-fiber echelle spectrograph with polarising capabilities. 
The dispersion for the spectropolarimeter is about 64\,000,
covering a wide spectral range, from 3700 to 10480~\AA\ with only a few small gaps in the near-infrared.
We selected two reddened targets, HD\,197770 and HD\,194279, for an in-depth study of the interstellar 
line polarisation. A summary of line-of-sight properties is provided in Table~\ref{tb:targets}. 
Exposure times amounted to 3040 and 5840 seconds for each final Stokes $Q$, $U$ and $V$ spectrum (each consisting of four sub-exposures taking at
different positions of the retarder) for HD\,197770 and HD\,194279, respectively.

The observations were automatically reduced with Upena, which is CFHT's reduction pipeline for ESPaDOnS. 
The Upena data reduction system uses Libre-ESpRIT which is a purpose built data reduction software tool (\citealt{1997MNRAS.291..658D}).
We choose to opt for continuum normalized spectra (we are looking for line variation) and to apply both the heliocentric velocity correction
and the radial velocity correction from telluric lines.
Thus we obtain total intensity stokes $I$ spectra as well as Stokes $Q/I$, and $U/I$ (linear), and $V/I$ (circular) spectra normalized to a zero mean
(\emph{i.e.}\ the spectra are sensitive to line polarisation only).
The achieved signal-to-noise ratio (S/N) varies across the spectrum, but is about 700 and 1200 in the red range, for HD\,194279 and HD\,19770, respectively.
In addition to the shape of the SED, the reduced S/N at the longest wavelengths is also due to the lower efficiency of the instrument 
(mainly the detector) and at the shortest wavelengths due to stronger extinction by dust.

\section{Results and discussion}

\subsection{Polarisation and total intensity line profiles}

The polarisation efficiency of an absorption line can be written generally as 
(adapted from \citealt{1974psns.coll..916G}, \citealt{1974ApJ...188..517M}, \citealt{1996ASPC...97..143S}):
\begin{equation}
\Delta P(\lambda) = f_{P}(\lambda)\ P(\lambda)\ \Delta\tau(\lambda) / \tau(\lambda)
\end{equation}
where $\tau(\lambda)$ and $P(\lambda)$ are the continuum optical depth and the continuum polarisation.
$\Delta\tau(\lambda)$ (\emph{i.e.} ${\rm ln}\,(I_{\rm c} / I_\lambda)$) 
is the observed change in optical depth across the line profile
and $f_{P}(\lambda)$ is the polarisation efficiency factor across the line profile.
$P(\lambda)$ can be computed from Eq.~\ref{eq:serkowski} if $P_{\rm max}$ and $K$ are known.
Therefore, with $\tau_\lambda = A_\lambda/1.086$ we can write the previous as
\begin{equation}
\Delta P(\lambda) = f_{P}(\lambda)\ \Delta\tau(\lambda) \ 1.086\ P(\lambda)/A(\lambda)\label{eq:DeltaP}
\end{equation}
where $f_{P}(\lambda)$ is the unknown polarisation efficiency parameter and $A(\lambda)$ is the extinction curve which
depends on $A_V$ and $R_V$ (see \emph{e.g.} \citealt{1989ApJ...345..245C}, 
\citealt{1994ApJ...422..158O}, \citealt{1999PASP..111...63F}, \citealt{2007ApJ...663..320F}).
\citet{2008JQSRT.109.1527V} derive $P(\lambda)/A(\lambda) \propto \lambda^\epsilon$, with $\epsilon = 1.41$ for 
HD\,197770 (from 1000~\AA\ to 1 $\mu$m), which matches with the power-law index of 1.4 computed for $P(\lambda)/A(\lambda)$
using Eq.~\ref{eq:DeltaP} and the $R_V = 2.8$ extinction curve (see \citealt{2007ApJ...663..320F}).
Applying the same procedure to HD\,194279 gives $\epsilon \sim 1.32$ in the optical.
Thus, the expected polarisation signal $\Delta P$ for a DIB depends on its wavelength.
For example, compared to the red DIBs near 5500~\AA\  ({\emph e.g} V band) the near-infrared 
DIBs ($\sim$ 7500~\AA) are then predicted, for the same $f_P$, to give rise to a polarisation 
signal a factor $\sim$1.5 stronger.
For grain related polarisation the efficiency is predicted to be a constant for a given transition,
thus the polarisation profile would have the same shape as the optical depth profile.
\citet{1974ApJ...188..517M} estimated for grain related carrier that $f_P \approx$ 1.0 -- 1.8.
We can also write: $f_P = \frac{\Delta P}{\Delta \tau_\lambda}\ \frac{\tau}{P}$.

However, we have no a priori information on the polarisability efficiencies of the DIB
carriers, which would scale with the amount of material (\emph{i.e.} column density)
and could be much higher for carriers of the weak DIBs.
The measured equivalent width (or central depth) is proportional to both column density $N$ and oscillator strength $f$.
For example, a weak DIB could be a result of a small oscillator strength, $f$ (or small abundance), of the particular
electronic transition, while the polarisation efficiency, $f_P$ could be large relative to the line strength (or abundance).
Or vice versa, a strong DIB could be due to a carrier with a large oscillator strength but with a small 
polarisation efficiency; \emph{i.e.}\ the ratio $f_P/f$ is not known for any of the DIB carrier candidates.
Thus, we stress the importance of investigating both strong and weak bands for polarisation features. 

\begin{figure*}[!ht]
 \centering
  \includegraphics[angle=0,width=14cm,clip]{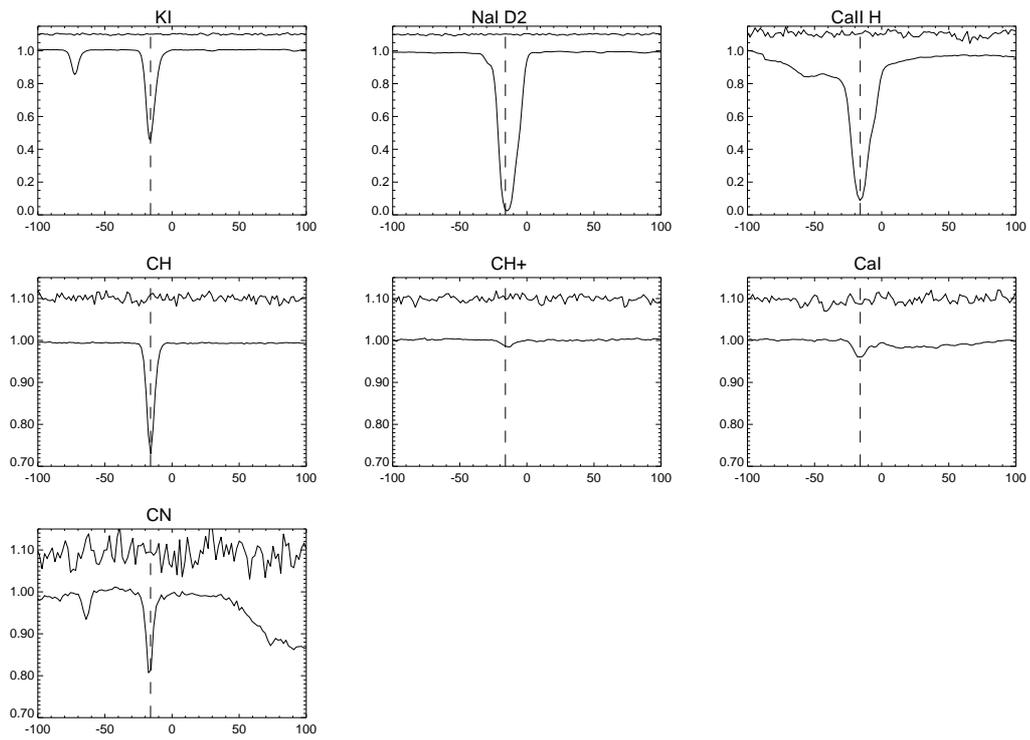}
   \caption{Normalised $I$ and $P$ (5x) spectra of atomic and molecular transitions as a function of 
   heliocentric velocity (in km\,s$^{-1}$) toward HD\,197770.
   From top left to bottom right \ion{Na}{i}~D, \ion{K}{i}, CH, CH$+$, \ion{Ca}{i}, \ion{Ca}{ii}, and CN.
   The dashed vertical lines indicate the interstellar radial velocity.}
   \label{fig:atommol_hd197770}
\end{figure*}

\begin{figure*}[!ht]
 \centering
  \includegraphics[angle=0,width=14cm,clip]{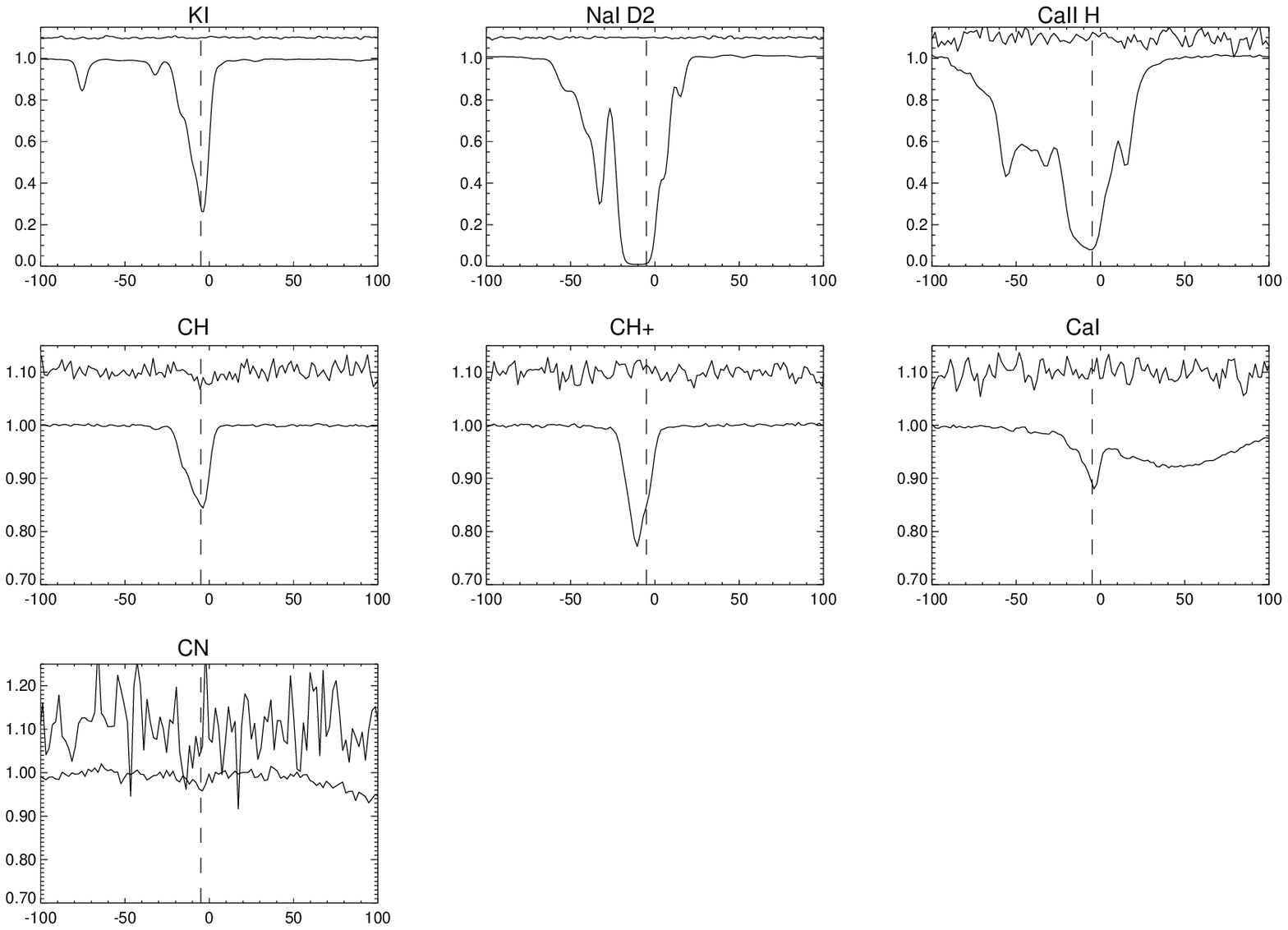}
   \caption{Normalised $I$ and $P$ (5x) spectra of atomic and molecular transitions as a function of 
   heliocentric velocity (in km\,s$^{-1}$) toward HD\,194279.
   From top left to bottom right \ion{Na}{i}~D, \ion{K}{i}, CH, CH$+$, \ion{Ca}{i}, \ion{Ca}{ii}, and CN.
   The dashed vertical lines indicate the interstellar radial velocity.}
   \label{fig:atommol_hd194279}
\end{figure*}

Intensity and polarisation spectra of CH, CH$^+$, CN, C$_2$, C$_3$, \ion{Na}{i}, \ion{Ca}{i}, \ion{Ca}{ii}, \ion{K}{i}
are shown in Figures~\ref{fig:atommol_hd197770} and~\ref{fig:atommol_hd194279} (Online).
Measurement of interstellar line strengths are given in Table~\ref{tb:is-lines}.
The polarisation spectra of the (di)atomic absorption lines do not reveal any line polarisation.

\begin{figure*}[ht!]
 \centering
  \includegraphics[angle=0,width=18cm,clip]{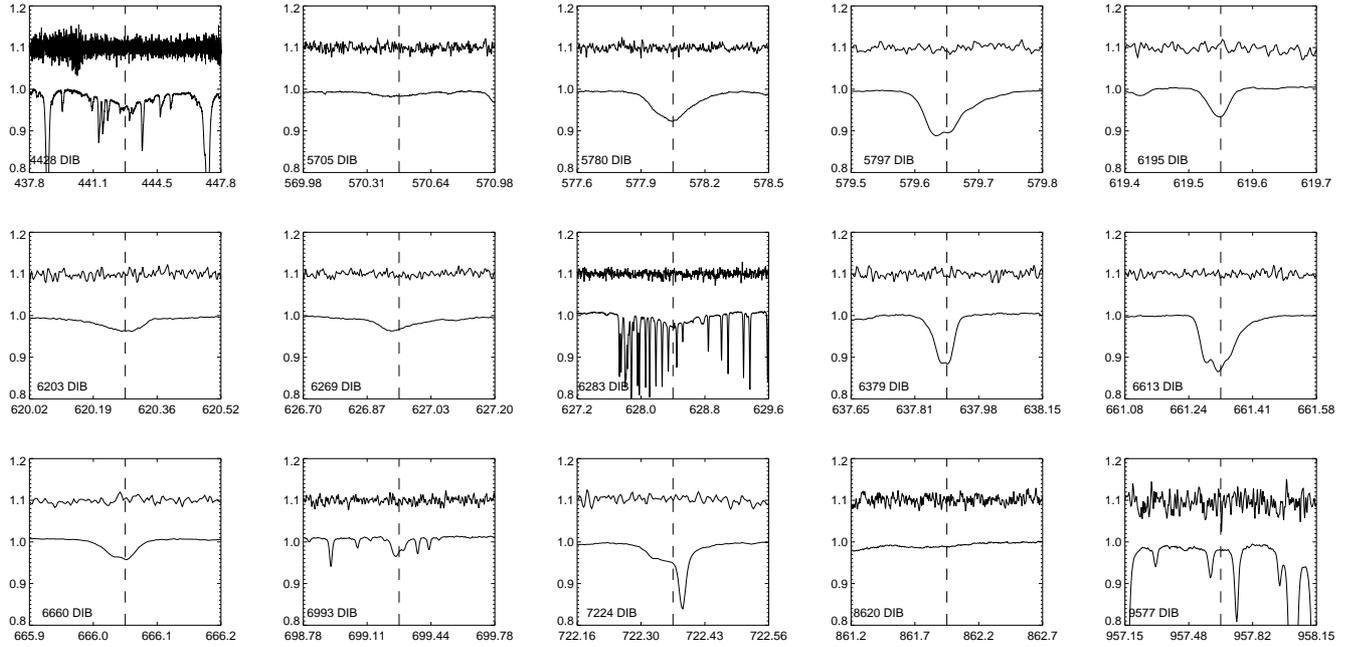}
   \caption{Total normalized intensity $I$ and polarisation $\Delta P$  spectra of 15 strongest DIBs toward HD\,197770.
   The $\Delta P$ spectra are scaled 10x and displaced vertically for display.
   DIB rest wavelengths from \citet{2008ApJ...680.1256H}, corrected for radial velocity of \ion{K}{i}, are shown as dashed vertical lines.   }
   \label{fig:spectraIP_hd197770_strongest}
\end{figure*}

\begin{figure*}[ht!]
 \centering
  \includegraphics[angle=0,width=18cm,clip]{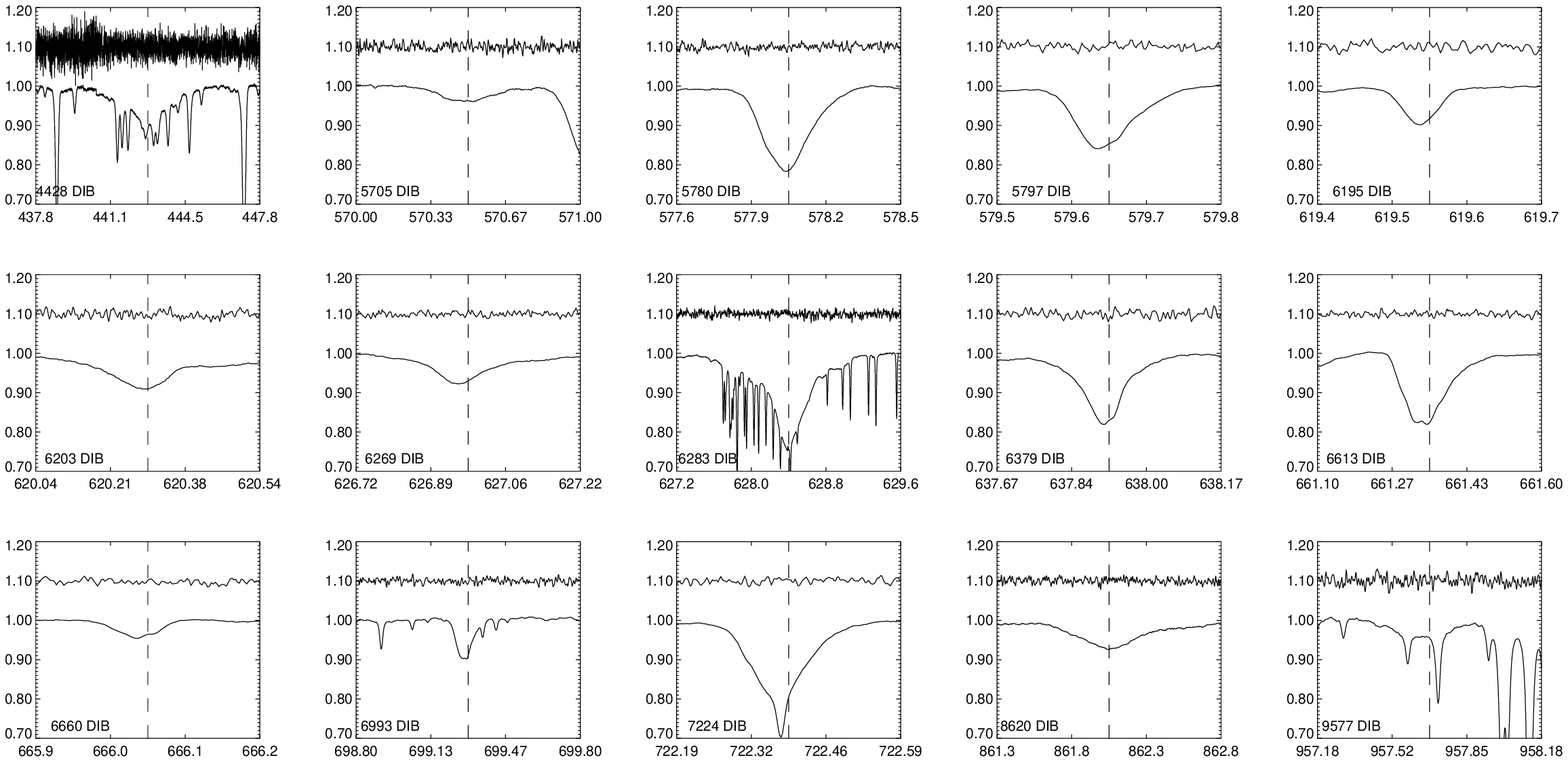}
   \caption{Total normalized intensity $I$ and polarisation $\Delta P$ spectra of 15 strongest DIBs toward HD\,194279.
   The $\Delta P$ spectra are scaled 10x and displaced vertically for display.
   DIB rest wavelengths from \citet{2008ApJ...680.1256H}, corrected for radial velocity of \ion{K}{i}, are shown as dashed vertical lines.    }
   \label{fig:spectraIP_hd194279_strongest}
\end{figure*}

\begin{figure*}[ht!]
 \centering
  \includegraphics[angle=90,width=11cm,clip]{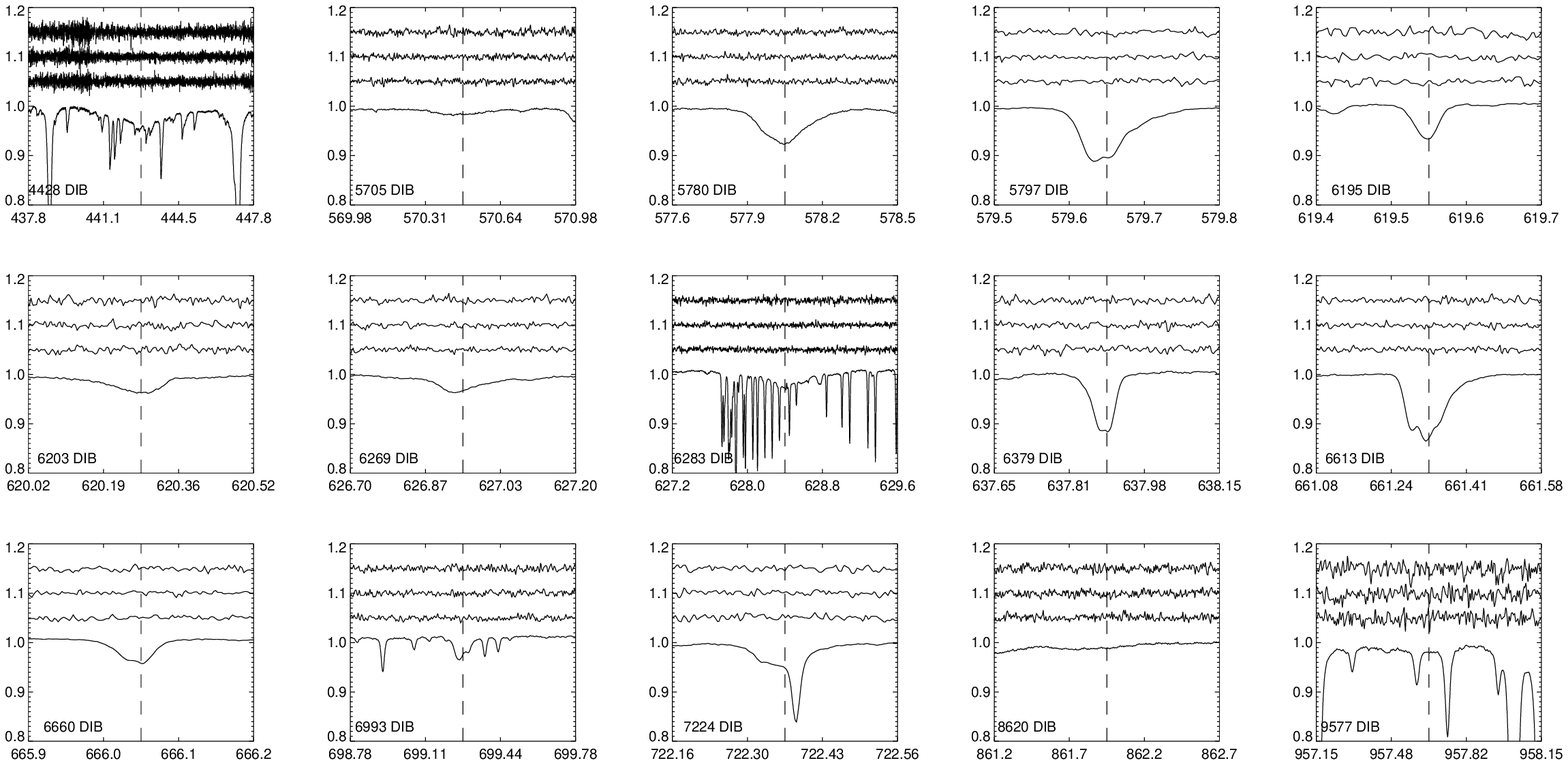}
   \caption{Stokes $V$, $U$, $Q$ and $I$ spectra (top to bottom) of 12 strong DIBs toward HD\,197770.
   The $Q, U, V$ spectra are scaled 5x and displaced vertically for display.}
   \label{fig:4stokes_hd197770_strongest}
\end{figure*}

\begin{figure*}[ht!]
 \centering
    \includegraphics[angle=90,width=13cm,clip]{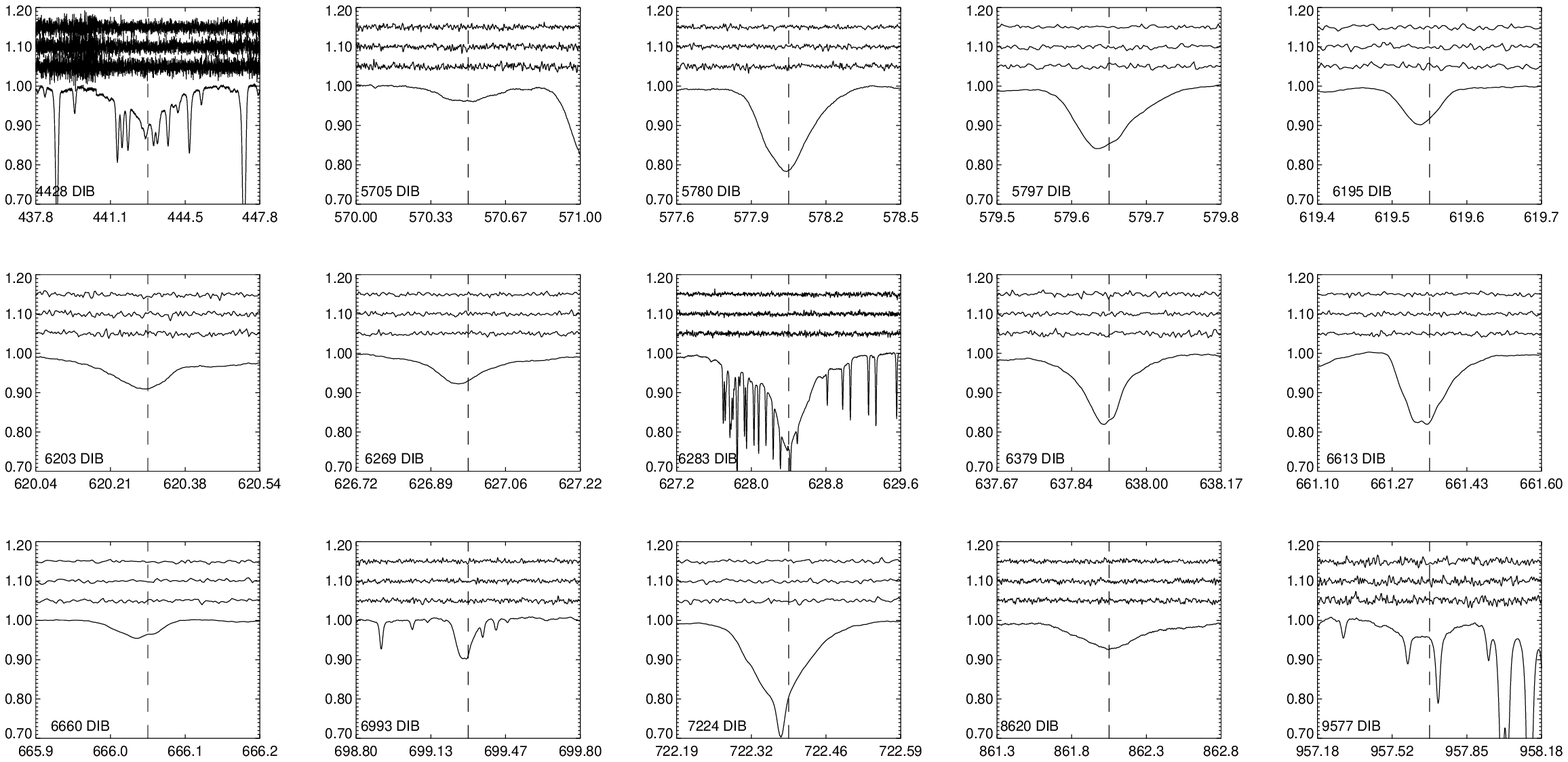}
   \caption{Stokes $V$, $U$, $Q$ and $I$ spectra (top to bottom) of 12 strong DIBs toward HD\,194279.
   The $Q, U, V$ spectra are scaled 5x and displaced vertically for display.  }
   \label{fig:4stokes_hd194279_strongest}
\end{figure*}

To exploit the large spectral range and high quality of the spectra obtained it is possible to explore the line
polarisation of not just the strongest DIBs but also those of moderate and weak strength, at both optical and near-infrared wavelengths.
First we focus on the strongest known diffuse bands. From the survey of HD\,183143 by \citet{1995ARA&A..33...19H} we select
all DIBs with central depths larger than 7\%, however we omit some of the very broad bands (\emph{i.e.}
at 5778, 6177, and 6533~\AA) as these are too difficult to detect in our echelle spectra.
This cut-off is somewhat arbitrary but at least ensures that all well-studied DIBs are included (Table~\ref{tb:poleff}).
Figures~\ref{fig:spectraIP_hd197770_strongest} and~\ref{fig:spectraIP_hd194279_strongest} show the observed line polarisation computed as follows:
\begin{equation} 
\Delta P = \sqrt{\frac{1}{3} [(P_{\rm m}+Q/I)^2 + (P_{\rm m}+U/I)^2 + (P_{\rm m}+V/I)^2]} - P_{\rm m}
\end{equation}
for the 15 strong DIBs at $\lambda\lambda$4428, 5705, 5780, 5797, 6196, 6203, 6269, 6283, 6379, 6613, 6660, 6993, 7224, 8621, and 9577 toward 
HD\,197770 and HD\,194279. Individual $V$, $Q$ and $U$ spectra are shown in Figures~\ref{fig:4stokes_hd197770_strongest} 
and ~\ref{fig:4stokes_hd194279_strongest} (Online).

\begin{figure*}[!ht]
 \centering
  \includegraphics[angle=90,width=15cm,clip]{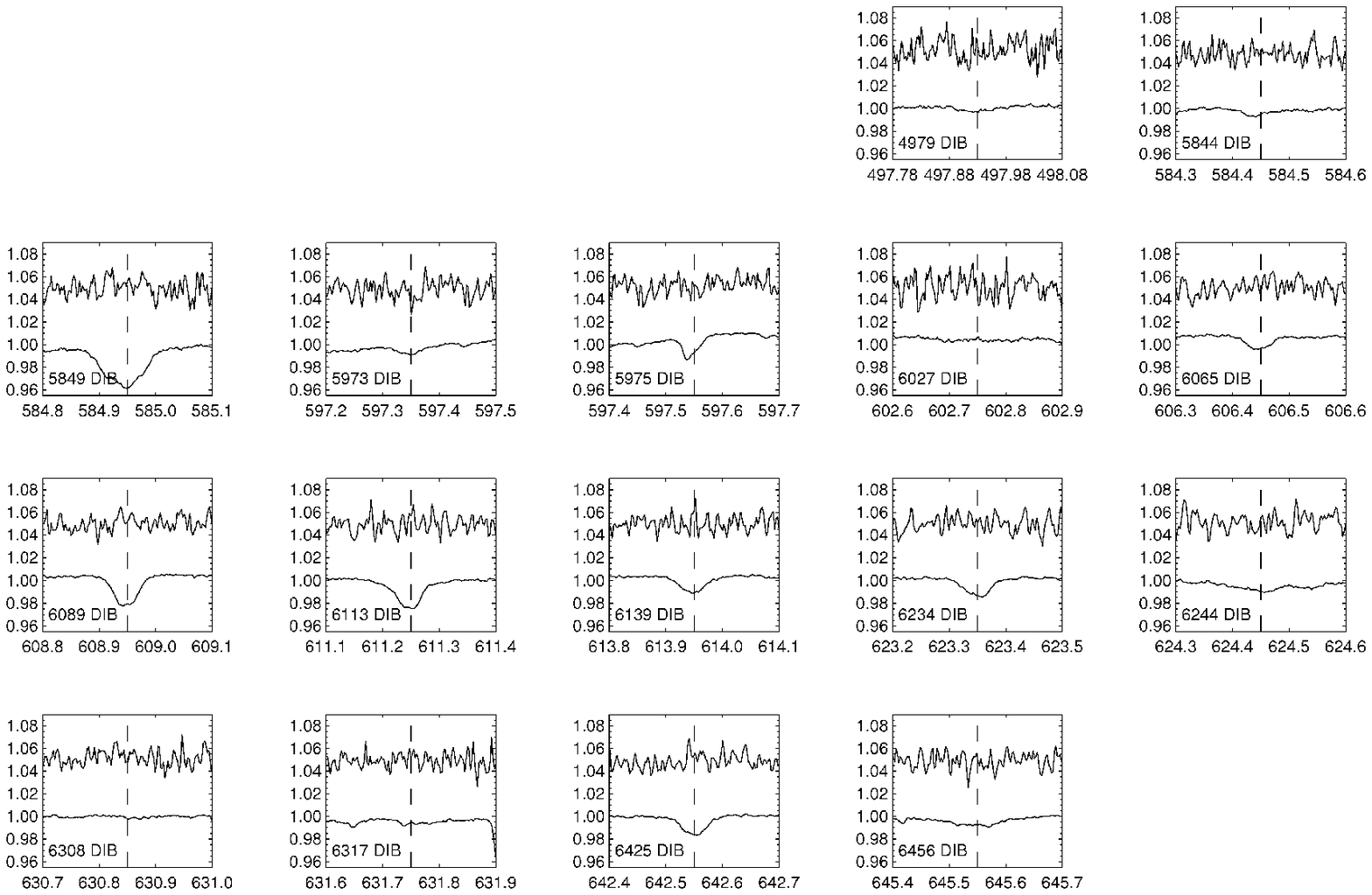}
   \caption{$\Delta P$ (top) and normalised intensity (bottom) spectra of 16 weak DIBs toward HD\,197770.
   The $\Delta P$ spectra are scaled 10x and displaced vertically for display.}
   \label{fig:spectraIP_weak_hd197770}
\end{figure*}

\begin{figure*}[!ht]
 \centering
  \includegraphics[angle=90,width=16cm,clip]{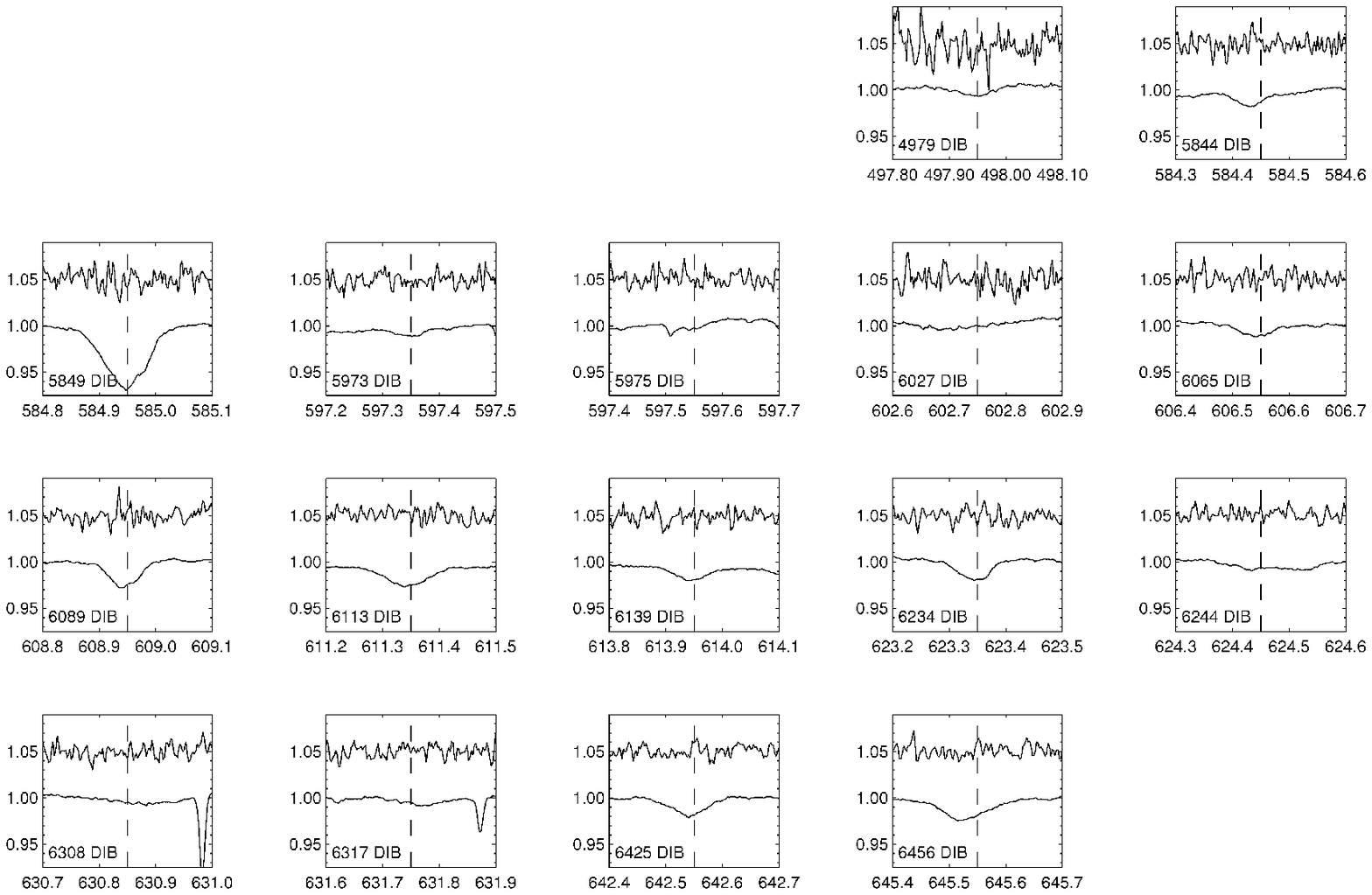}
   \caption{$\Delta P$ (top) and normalised intensity (bottom) spectra of 16 weak DIBs toward HD\,194279.
   The $\Delta P$ spectra are scaled 10x and displaced vertically for display.  }
   \label{fig:spectraIP_weak_hd194279}
\end{figure*}

\begin{figure*}[!ht]
 \centering
   \includegraphics[angle=90,width=16cm,clip]{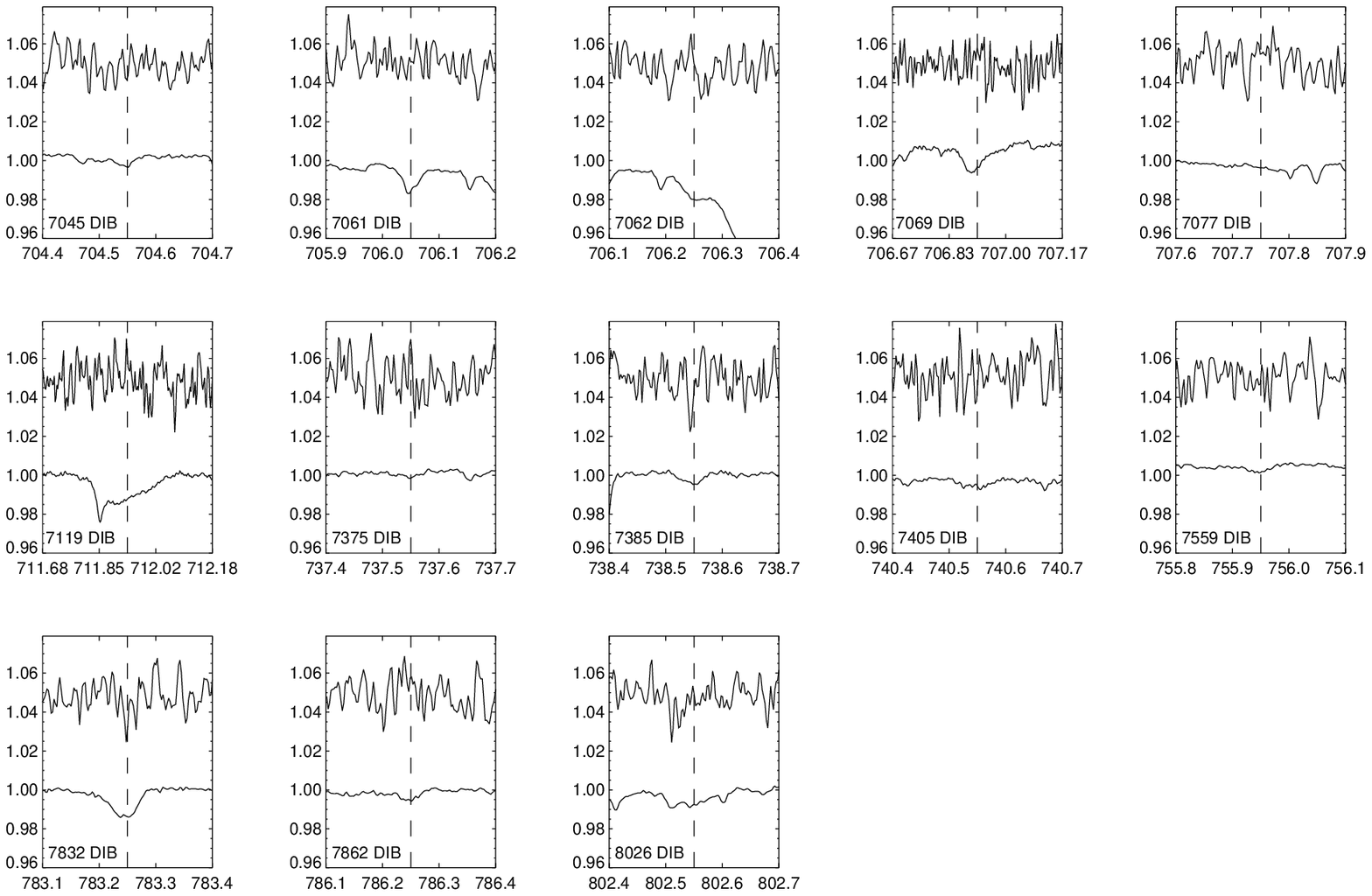}
  \caption{$\Delta P$ (top) and normalised intensity (bottom) spectra of 13 near-infrared DIBs toward HD\,197770.
   The $\Delta P$ spectra are scaled 10x and displaced vertically for display.}
   \label{fig:spectraIP_nir_hd197770}
\end{figure*}

\begin{figure*}[!ht]
 \centering
   \includegraphics[angle=90,width=16cm,clip]{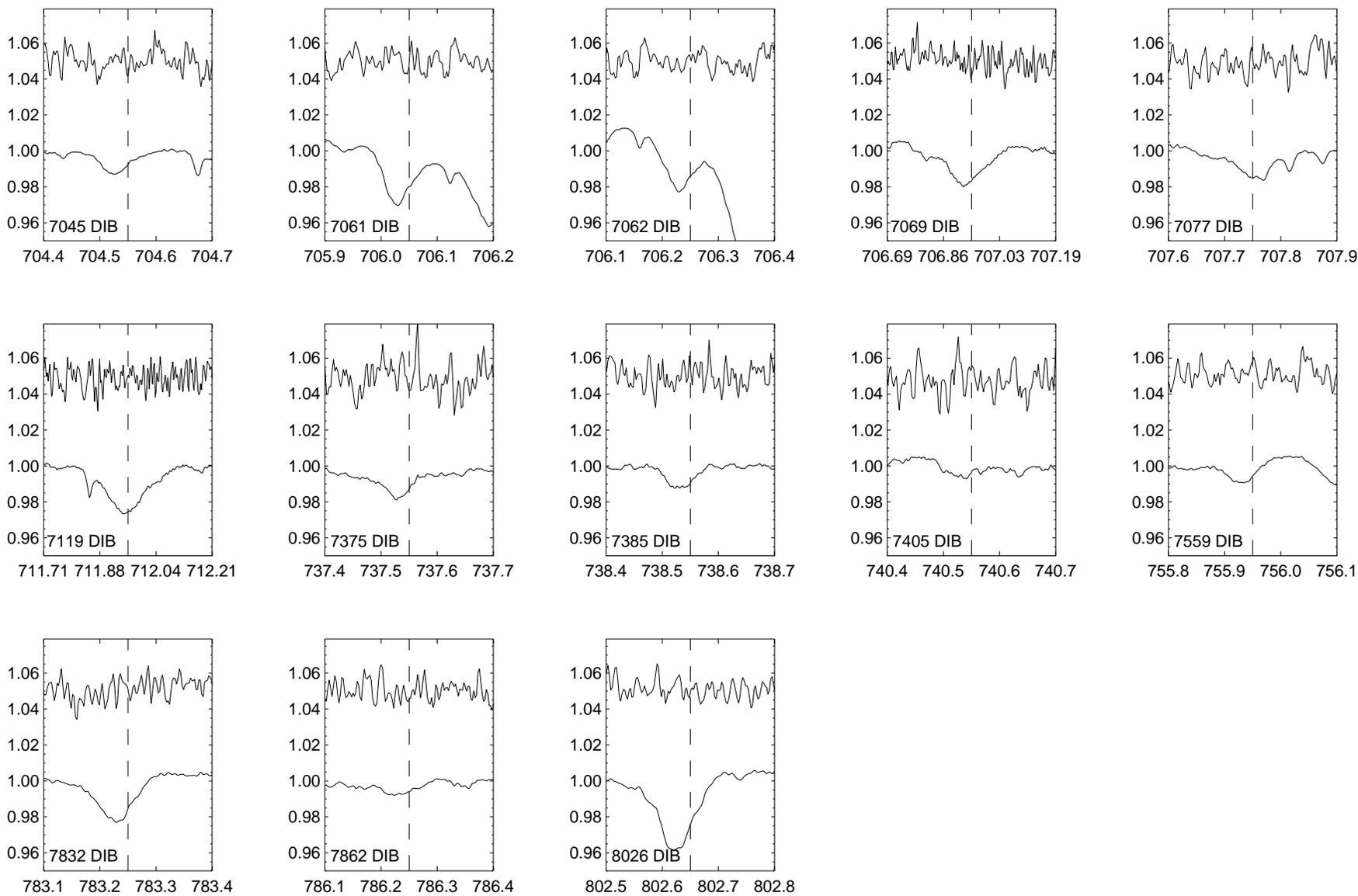}
  \caption{$\Delta P$ (top) and normalised intensity (bottom) spectra of 13 near-infrared DIBs toward HD\,194279.
   The $\Delta P$ spectra are scaled 10x and displaced vertically for display.  }
   \label{fig:spectraIP_nir_hd194279}
\end{figure*}

\begin{figure*}[!ht]
 \centering
    \includegraphics[angle=90,width=17cm,clip]{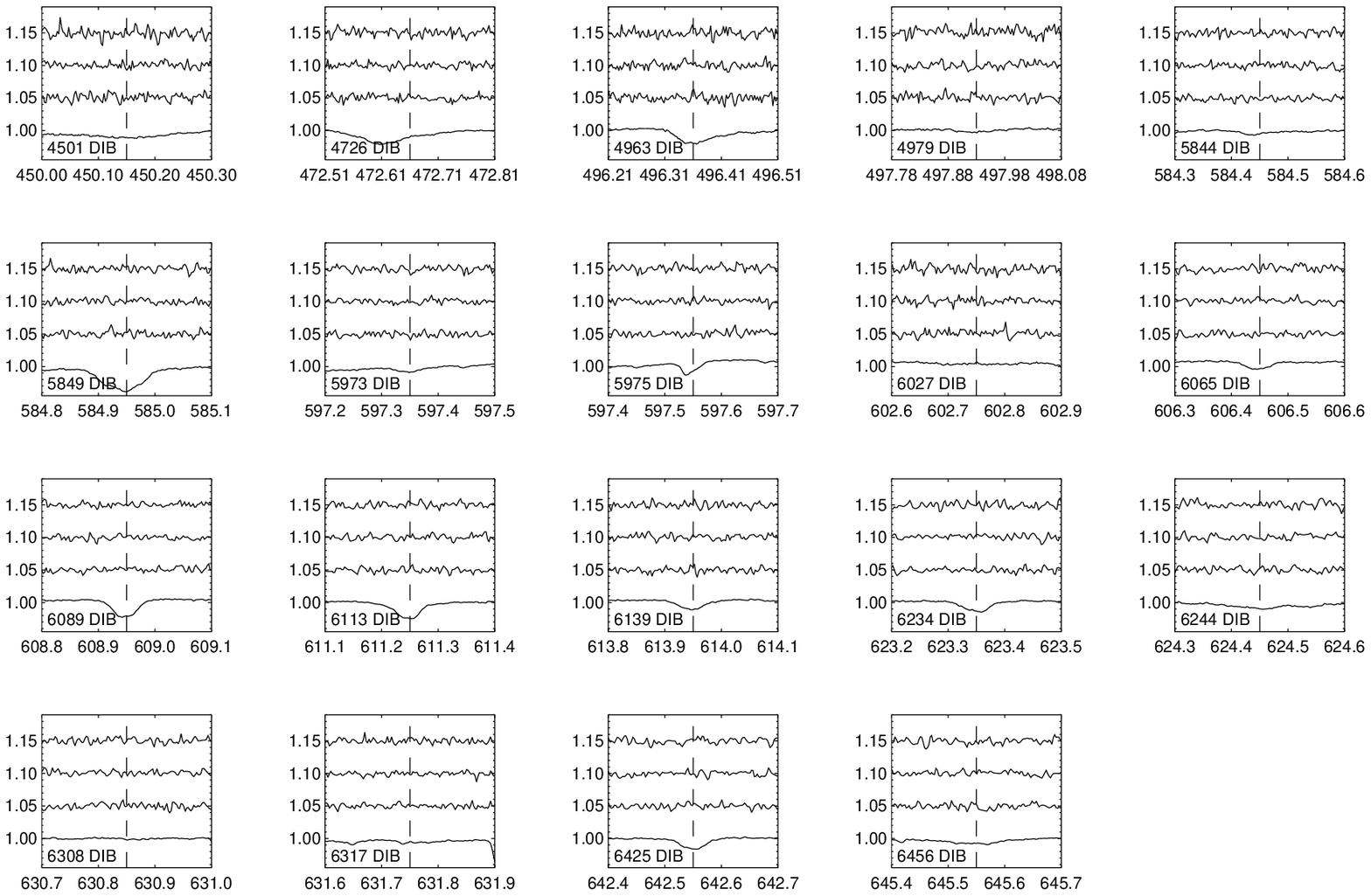}
  \caption{Stokes $V$, $U$, $Q$ and $I$ spectra (top to bottom) of 16 weak DIBs toward HD\,197770.
   The $Q, U, V$ spectra are scaled 5x and displaced vertically for display.}
   \label{fig:4stokes_weak_hd197770}
\end{figure*}

\begin{figure*}[!h]
 \centering
    \includegraphics[angle=90,width=17cm,clip]{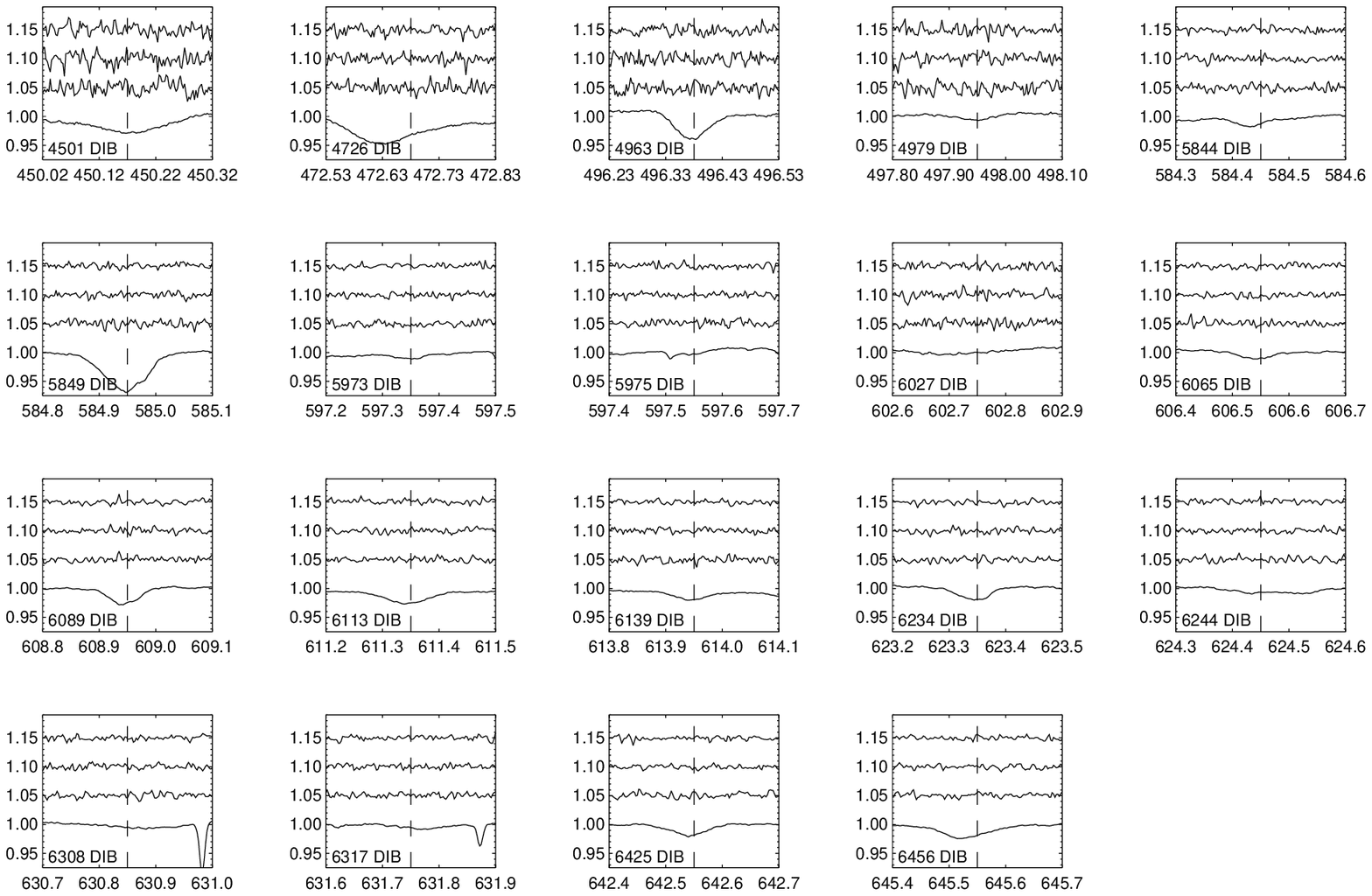}
   \caption{Stokes $V$, $U$, $Q$ and $I$ spectra (top to bottom) of 16 weak DIBs toward HD\,194279.
   The $Q, U, V$ spectra are scaled 5x and displaced vertically for display.  }
   \label{fig:4stokes_weak_hd194279}
\end{figure*}

\begin{figure*}[!ht]
 \centering
    \includegraphics[angle=90,width=17cm,clip]{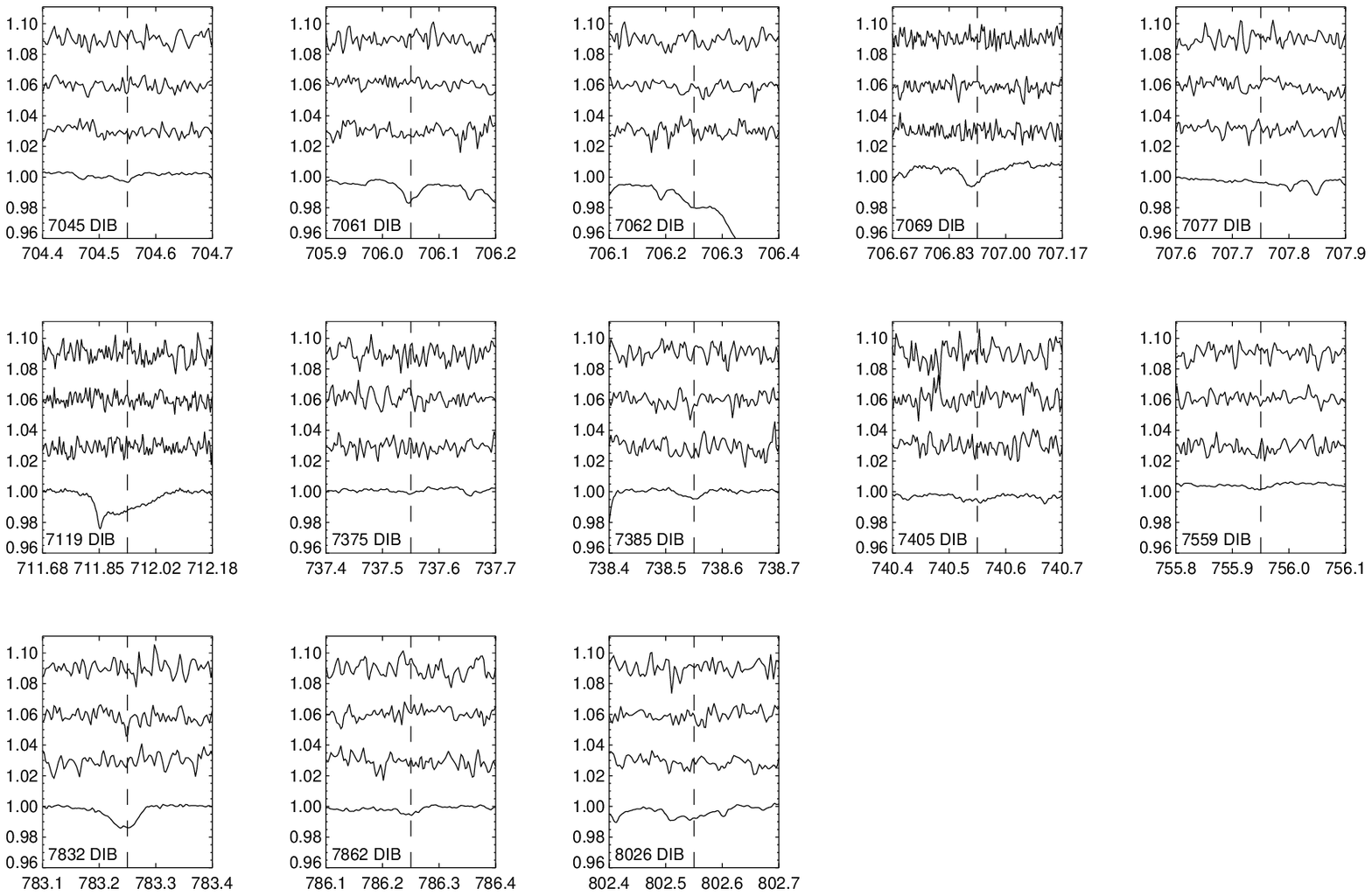}
 \caption{Stokes $V$, $U$, $Q$ and $I$ spectra (top to bottom) of 13 near-infrared DIBs toward HD\,197770.
   The $Q, U, V$ spectra are scaled 5x and displaced vertically for display.}
   \label{fig:4stokes_nir_hd197770}
\end{figure*}

\begin{figure*}[!h]
 \centering
    \includegraphics[angle=90,width=17cm,clip]{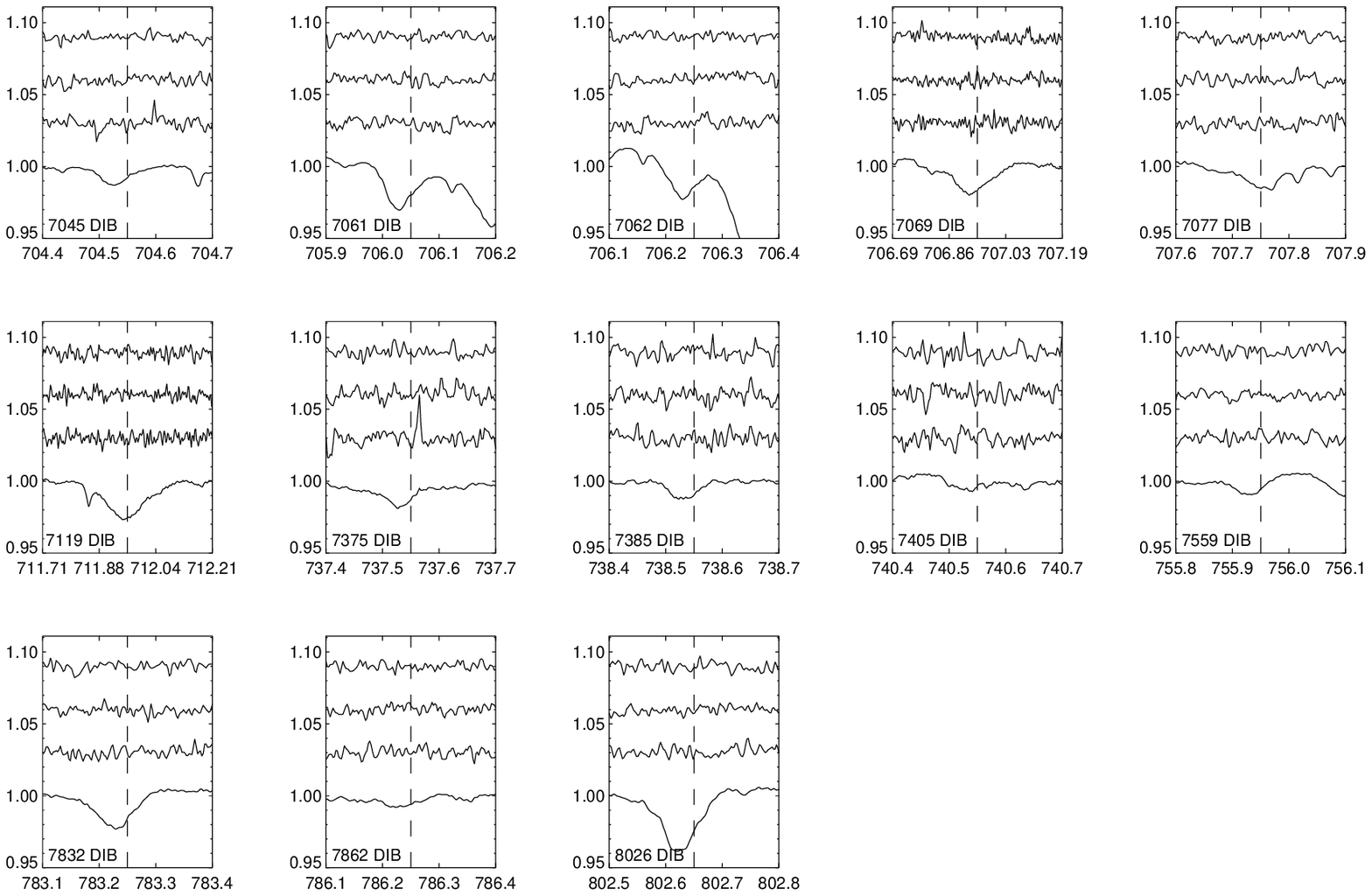}
  \caption{Stokes $V$, $U$, $Q$ and $I$ spectra (top to bottom) of 13 near-infrared DIBs toward HD\,194279.
   The $Q, U, V$ spectra are scaled 5x and displaced vertically for display.  }
   \label{fig:4stokes_nir_hd194279}
\end{figure*}

Complementary to previous polarisation studies and to account for the possible strong alignment / polarisation signal
of weak(er) DIB carriers (see previous section) we include also a selection of these in this work.
We select sixteen DIBs of moderate and weak strength confirmed by several previous DIB surveys (\citealt{1994A&AS..106...39J},
\citealt{2000A&AS..142..225T}, \citealt{2009ApJ...705...32H}).  In addition, we selected also 13 near-infrared DIBs for analysis.
In our selection we avoided DIBs that might be contaminated by either stellar or telluric lines, strong adjacent DIBs, or those that are too weak 
to be detected in the most reddened sightline towards HD\,194279.
See Table~\ref{tb:poleff} and Figs.~\ref{fig:spectraIP_weak_hd197770} to~\ref{fig:spectraIP_nir_hd194279} for the list of DIBs and
the corresponding intensity and polarisation $\Delta P$ spectra.
Again, the corresponding $Q$, $U$, and $V$ spectra are shown in Figs.~\ref{fig:4stokes_weak_hd197770} to~\ref{fig:4stokes_nir_hd194279} (Online).

\begin{table*}[ht!]
\caption{Upper limits on the DIB carrier polarisation efficiency factor $f_P$. See text for details.}
\label{tb:poleff}
\centering
\resizebox{16cm}{!}{
\begin{tabular}{lll|lll|lll}\hline\hline
\multicolumn{9}{c}{Upper limits on the polarisation efficiency factor $f_P$} \\ \hline   
DIB\tablefootmark{a}	   &  HD\,197770  &  HD\,194279 &  DIB       &  HD\,197770 &  HD\,194279   &  DIB      &  HD\,197770 &  HD\,194279 \\ \hline
4428.19	   &$\sim$0.033   &$\sim$0.070  &  4979.61   &	0.15       & 0.39          &  7045.89  & 0.06	     & 0.11        \\
5705.08	   &  0.056	  &  0.068	&  5844.92   &  0.09 	   & 0.16	   &  7061.05  & 0.03	     & 0.05	   \\
5780.48	   &  0.007	  &  0.008	&  5849.81   &  0.02 	   & 0.03 	   &  7062.68  & 0.07	     & 0.07	   \\
5797.06	   &  0.005	  &  0.010	&  5973.81   &  0.08 	   & 0.24	   &  7069.55  & 0.03	     & 0.07	   \\
6195.98    &  0.010	  &  0.024	&  5975.75   &  0.03  	   & -   	   &  7077.86  & -	     & 0.14	   \\
6203.05	   &  0.019	  &  0.024	&  6027.68   &  -   	   & 0.24	   &  7119.71  & 0.03	     & 0.06	   \\
6269.85	   &  0.015	  &  0.017	&  6065.28   &  0.05 	   & 0.15	   &  7375.87  & - 	     & 0.17	   \\
6283.84	   &  0.017	  &  0.006	&  6089.85   &  0.02 	   & 0.07 	   &  7385.89  & 0.08	     & 0.14	   \\
6379.32	   &  0.005	  &  0.010	&  6113.18   &  0.02 	   & 0.08 	   &  7405.71  & 0.10	     & 0.25	   \\
6613.62	   &  0.003	  &  0.007	&  6139.98   &  0.04 	   & 0.13	   &  7559.41  & 0.08	     & 0.10	   \\
6660.71	   &  0.010	  &  0.033	&  6234.01   &  0.04 	   & 0.08 	   &  7832.89  & 0.02	     & 0.06	   \\
6993.13	   &  0.014	  &  0.015	&  6244.46   &  0.07 	   & 0.17	   &  7862.43  & 0.07	     & 0.23	   \\
7224.03	   &  0.013	  &  0.007	&  6308.80   &  -   	   & 0.14	   &  8026.25  & 0.04	     & 0.03	   \\
8620.41	   &  0.11	  &  0.029	&  6317.86   &  -    	   & 0.07 	   &           &	     &  	   \\
9577.00    &  -  	  &$\sim$0.095	&  6425.66   &  0.03 	   & 0.07 	   &           &	     &  	   \\
           &		  &		&  6456.01   &  0.07 	   & 0.06 	   &           &	     &  	   \\
\hline
\end{tabular}									      
}
\tablefoot{
\tablefoottext{a}{DIB central wavelengths from \citet{2008ApJ...680.1256H} are with respect to \ion{K}{i}. 
Central wavelength for 8620 DIB from \mbox{\citet{2008A&A...488..969M}} and for 9577 DIB from Herbig (1995). 
For DIB non-detections the upper limits could not be derived (``-'').}
}
\end{table*}

\subsection{Environmental conditions of the sightlines}
In this section we review the physical conditions of the interstellar medium towards HD\,197770 and HD\,194279.

\subsubsection{HD\,197770} 
HD\,197770 is an evolved, spectroscopic, eclipsing binary with two B2 stars (\emph{e.g.}\ \citealt{1996PASP..108..401C}).
It lies on the edge of a large area of molecular clouds and star formation in the Cygnus region, at the edge of Cyg OB7 and Cep OB2
(\citealt{1998AJ....115.2561G}), at a distance of $\sim$440~pc (\citealt{1994ApJS...95..419D}).

The interstellar medium in front of HD\,197770 has also been studied extensively, in particular because it is currently 
one of two sightlines (the other being HD\,147933-4) for which a polarisation feature (at level of 0.4\% and efficiency 
$\Delta p$/ $\Delta\tau$ = 0.0017)
corresponding to the 2175~\AA\ UV bump has been detected (\citealt{1992ApJ...385L..53C}, \citealt{1994ApJ...427L..47S}, 
\citealt{1994ApJ...431..783K}, \citealt{1995ASSL..202..271M}, \citealt{1997ApJ...478..395W}) as predicted by \citet{1989IAUS..135..313D}.
\citet{1992ApJ...385L..53C}, \citet{1997ApJ...478..395W} assigned the polarisation of the UV bump to small aligned graphite disks,
while other authors favour silicate grains (\citealt{1994ApJ...431..783K}).
The sightline shows a high continuum linear polarisation of almost 4\% in the optical (see also Table~\ref{tb:targets}).

The dust grains in this interstellar cloud are aligned, where $P_{\rm max} / \tau_V$ is 0.026, 
which is close to optimal alignment (0.032; \citealt{1975ApJ...196..261S}).

From optical spectroscopy we observe strong CH and CN absorption lines, but a weak CH$^+$ line.
Column densities of \ion{Ca}{i}, \ion{Fe}{i}, CH, CH$^+$, CN and C$_2$ for the main velocity component at -3~km\,s$^{-1}$\
have been reported by \citet{1994ASPC...58...84H} and are consistent with our values (Table~\ref{tb:is-lines}).
The atomic and molecular line profiles show a single strong narrow component at a heliocentric velocity of -17~km\,s$^{-1}$.
The CN and CH lines are narrow, with FWHM of $\sim$0.07 to 0.09~\AA, while CH$^+$, \ion{Ca}{i}, and \ion{Ca}{ii} are
a little broader, with FWHM of $\sim$0.16 to 0.20~\AA.

Both the 5797 and 5780~\AA\ DIBs are weak, per unit reddening, with respect to the Galactic average.
However, the strength ratio of 5797 over 5780 is relatively high ($W(5797)/W(5780) = 0.58$), 
typical of a translucent ($\zeta$-type) diffuse cloud.
This is also indicated by the low CH$^+$/CH ratio of 0.11 (lower than 0.5 is typical for a quiescent medium, no shocks).
Previously, \citet{1993ApJ...403..722W} invoked quiescence for this sightline to efficiently align the UV bump grains.

In summary, this sightline probes a {\it single dense quiescent interstellar cloud}.

\subsubsection{HD\,194279}\label{sec:hd194279}
HD\,194279 (Cyg OB9) is associated to the NGC\,6910 cluster which has a distance between 1.7 -- 2.1~kpc 
(\citealt{1956PDAO...10..201U}, \citealt{1930LicOB..14..154T}).
\citet{2002ApJ...567..391M} report a spectroscopic parallax distance of 1.1~kpc, while \citet{1994Ap.....37..317G} derived
a distance of 740~pc for the Cyg\,OB9 association.
The sightline toward HD\,194279 shows a more complex structure with components at -16.1, -9 and -2.6~km s$^{-1}$.
It shows multiple strong components in both CH and CH$+$ whereas the CN line is very weak.
The CH$^+$, CH and CN lines have similar FWHM of 0.20, 0.22, and 0.17~\AA, respectively.
To explain high CH$^+$ abundances in the diffuse ISM both shocks and a strong UV field are invoked in cloud models (\citealt{1996A&A...307..271S}).
The $C_2$ transitions are detected, originating from the coldest component only.
It thus appears that HD\,194279 probes several diffuse clouds which in superposition cause significant reddening.

The dust grains in this interstellar cloud are not efficiently aligned as illustrated by a very low value for 
$P_{\rm max} / \tau_V = 0.008$, which is far from optimal alignment.

The $W(5797)/W(5780)$ ratio in this line-of-sight is 0.32, which is close to the average Galactic value of $\sim$0.26 (Vos et al. in preparation).
Also the W(CH$^+$)/W(CH) ratio, of 1.36, is intermediate, a sign of a slight enhanced production of CH$^+$ in this sightline.
From the line profiles of the atomic and molecular lines it is clear that this line-of-sight probes multiple diffuse cloud components,
for which the entire sightline gives average Galactic values for DIB strengths and molecular line ratios.

In summary, this sightline probes an {\it average of diffuse interstellar clouds}.

\subsection{DIB polarisation limits}\label{sec:discussion}
Previously, \citet{2007A&A...465..899C} provided linear polarisation 2$\sigma$ detection limits per FWHM of 0.04 -- 0.14\% for 
HD\,199770 for the $\lambda\lambda$5780, 5797, 6196, 6283, 6379, and 6283 DIBs. 
Corresponding polarisation detection limits per \AA\  (PDLA) are 0.06 to 0.19\%. 
Or alternatively, $f_{\rm max}$ are 0.31, 0.44, 0.45, 0.18, 0.47, and 0.68 resp.
Circular line polarisation for these DIBs gave 2$\sigma$ (per 0.1~\AA) limits of 1.0 -- 2.5\% for HD\,197770.

In this work we derive new upper limits on the polarisation efficiency $f_P$ (\emph{i.e.}\ similar to $f_{\rm max}$ in \citealt{2007A&A...465..899C})
for 45 strong, weak and near-infrared DIBs (see Table~\ref{tb:poleff}).
$f_P(\lambda_{\rm peak})$ is computed from Eq.~\ref{eq:DeltaP} adopting  
$P(\lambda)/A(\lambda) = P_{\rm max}/A_V$, $\Delta P(\lambda) = 2 \sigma_{\Delta P}$ 
(the standard deviation per pixel on $\Delta P$), and $\Delta\tau_{\rm peak} = {\rm ln}(1/(1 - {\rm central depth} )$.
The new constraints on the level of line polarisation are given in Table~\ref{tb:poleff}.

In this work we investigate two different types of interstellar clouds that show evidence of interstellar polarisation due to dust grains.
In neither of these two environments do we detect polarisation significant signals, with detection limits on $f_P$ of about 0.02 for the strongest
DIBs toward HD\,197770. For example, for the 8621 DIB we derive $f_P < 0.11$. The strength of this DIB is known to be 
strongly correlated to the amount of interstellar dust and has been thus suggested to be related 
more directly to grains than other DIBs (\citealt{2007PASP..119.1268W}, \citealt{2008A&A...488..969M}). 
The theoretical $f_P$ for a classical grain carrier is a  factor 10 higher than this new limit.
The 9577 DIB, attributed to the C$_{60}^+$ fullerene, also does not reveal line polarisation, with $f_P < 0.1$ (towards HD\,194279).
For the weak DIBs we obtain upper limits on $f_P$ of $\sim$0.02 to $\sim$0.2 towards both sightlines.
Again these limits suggest a non-grain related carrier.
The near-infrared DIBs suggest $f_P$ values between 0.02 to 0.10 for these bands (for the HD\,197770 sightline;
factor of two less stringent for the HD\,194279 sightline).
These low levels for the polarisation efficiency exclude typical ``classical'' dust grains as carriers, and thus
strongly reinforce the idea that all DIB carriers are gas-phase molecules.
In particular, polycyclic aromatic hydrocarbons (PAHs) and fullerenes are proposed as candidates for the DIB carriers (See recent assessment
by \citealt{1996ApJ...458..621S}).
The presence of these large molecules in the ISM has been confirmed from their mid-infrared emission features in various astrophysical 
environments (\citealt{2008IAUS..251..357S}, \citealt{2008ARA&A..46..289T}).

Recently, \citet{2009ApJ...698.1292S} investigated the polarised infrared emission by PAHs upon anisotropic irradiation by UV photons and the 
subsequent alignment of the angular momentum and the principal axis of the PAH molecule. Conservation of the angular moment and partial 
memory retention of UV photon source direction leads to partial polarisation (of not more than few \%) of the infrared emission. 
This is an extension of the notion put forward by 
\citet{1988prco.book..769L} that infrared emission features resulting from in-plane and out-of-plane modes should have orthogonal polarization directions.
Additionally, Tolkachev and collaborators have shown from theoretical and experimental work that large molecules can show polarisation
signatures in fluorescent emission excitation lines (\emph{e.g.} \citealt{1994OptSp..77...38T}, \citealt{2009JApSp..76..806T}).
In the case of PAHs, it would be interesting to quantitatively assess the polarization efficiency that is associated with the electronic 
absorption of these molecules and their ions when aligned in an external magnetic field and compare this value to the values of $f_P$ derived from the observations. 
Recent advances in quantum chemical calculations of PAH polarizability (\citealt{2007JChPh.127a4107M}) should help make it possible 
to quantify the polarization that is associated with a population of PAH molecules or ions. Studies are ongoing in this direction and will be reported in a separate report. 

\section{Conclusion}

The results presented in this study show that:
\begin{enumerate}
\item The Polarisation Detection Limit per \AA\  in \% (\emph{i.e.}\ PDLA = 2$\sigma_{\Delta P}$(per pixel) / $\sqrt{1./ {\rm pixel size (\AA)}}\times 100.0$) 
for the DIBs in the red and green spectral range (\emph{i.e.} 
between 5700 and 7000~\AA) have typical values between 0.004 and 0.010\%, an order of magnitude improvement with
respect to previous limits.

\item None of the 45 DIBs measured and analysed in this work show unambiguous evidence for line polarisation of the DIBs.

\item For the strongest DIBs towards HD\,197770 the obtained upper limits on the polarisation efficiency factor are at least a 
factor 10 smaller (and in some cases more than 300 times smaller) than those expected for classical grains.

\item For all 45 DIBs the derived $f_P$ is significantly less than 1, the lower limit predicted for carriers related to classical grains.

\end{enumerate}

In summary, none of the diffuse bands of varying strengths and widths exhibit a polarised absorption spectrum,
neither for the dense cloud, with efficient grain alignment, in the line-of-sight towards HD\,197770, nor
the diffuse clouds averaged along in the line-of-sight toward HD\,194279.
Thus, we postulate it is likely that none of the DIB carriers measured in our study are directly related to grain-like carriers. 
This includes the 8621~\AA\ DIB for which a very good correlation with dust reddening has been observed.
Also, if DIB carriers are indeed related to large gas-phase molecules, it appears that these do indeed not align efficiently in 
the diffuse ISM and/or have a low polarisation efficiency.

\begin{acknowledgements}
We thank the CFHT queued service mode observers for help in preparing and executing the observations and help with the subsequent 
processing of the spectral data. 
P. Ehrenfreund is supported by NASA Grant NNX08AG78G and the NASA Astrobiology Institute (NAI).
This research has made use of the SIMBAD database, operated at CDS, Strasbourg, France.
We are thankful to the IDL Astronomy Library maintained at the Goddard Space Flight Center \citep{1993ASPC...52..246L}.
\end{acknowledgements}

\bibliographystyle{aa}  
\bibliography{/lhome/nick/Desktop/ReadingMaterial/Astronomy/Bibtex/bibtex} 

\Online

\onecolumn

\begin{appendix}
\section{Polarisation ($\Delta P$) and total intensity $I$ spectra for weak and near-infrared DIBs}

\end{appendix}

\clearpage

\begin{appendix} 

\section{Atomic, molecular and DIB line profiles, equivalent widths and central depths}

\begin{figure*}[!ht]
 \centering
   \includegraphics[bb=45 75 555 775,angle=-90,width=14cm,clip]{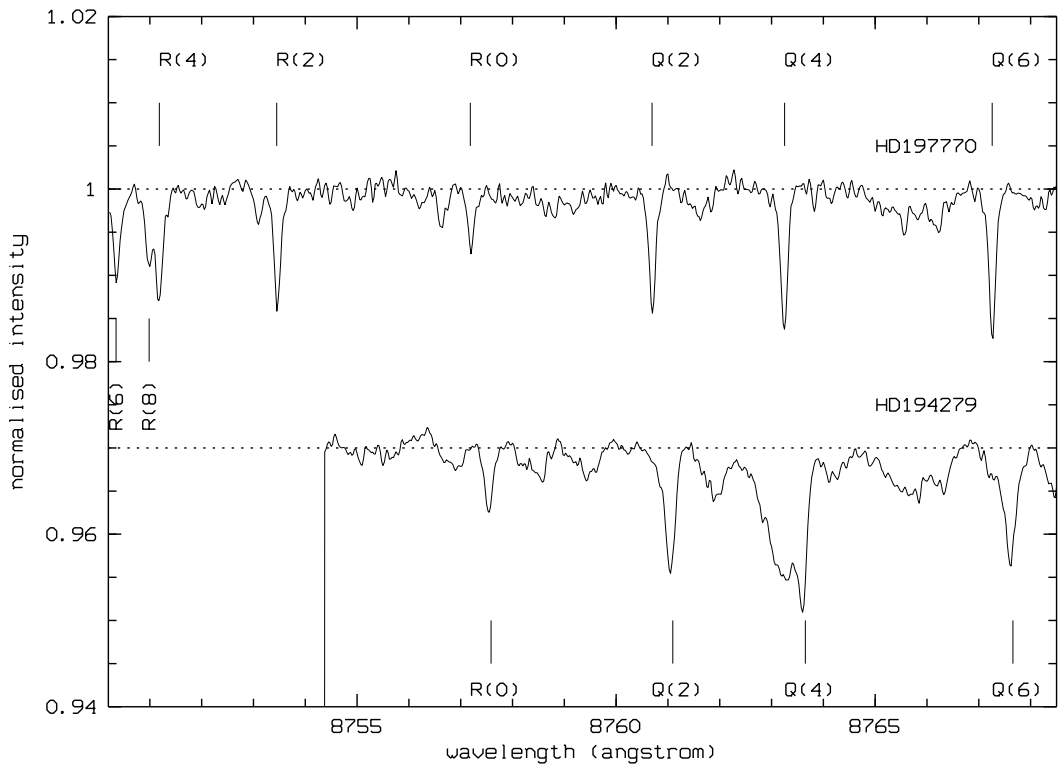}
  \caption{$C_2$ (2-0) Phillips band toward both sightlines. Line strengths are given in Table~\ref{tb:is-lines}.
   For HD\,194279 the strong Hydrogen Paschen line at 8750.47~\AA\ has been removed (no C$_2$ lines detectable),
   and furthermore, the 8764.4~\AA\ DIB is clearly present in this line-of-sight.}
   \label{fig:spectraC2}
\end{figure*}

\begin{table}[ht!]
\caption{Equivalent widths (m\AA) for interstellar atomic and diatomic lines observed for the two lines-of-sight.}
\label{tb:di-atomics}
\centering
\begin{tabular}{lll}\hline\hline
		& \object{HD 197770}		& \object{HD 194279}\tablefootmark{a}		\\ \hline 
CN (3874.6~\AA) & 20.0 $\pm$ 0.3		& 4.9 $\pm$ 0.2			\\
CN (3873.8~\AA)	& 4.1 $\pm$ 0.2			& -				\\
\ion{Fe}{i}(3719.9+3859.9)& 3.7$\pm$0.8,2.8$\pm$0.5& 9.6$\pm$4, 8.2$\pm$1       \\
\ion{Ca}{ii}\,K	& 186 $\pm$ 1			& 720 $\pm$ 1.5 		\\
\ion{Ca}{ii}\,H	& 96.5 $\pm$ 0.5		& 397 $\pm$ 3   		\\
CH$^+$ 4232.3~\AA& 3.0 $\pm$ 0.3		& 46.8 $\pm$ 0.3		\\
\ion{Ca}{i}	& 8.0 $\pm$ 0.3			& 11.9 $\pm$ 0.7		\\
CH (4300.3~\AA)	& 26.6 $\pm$ 0.1		& 34.4 $\pm$ 0.3		\\
\ion{K}{i} (4044~\AA)& 1.0 $\pm$ 0.1		& 1.0          			\\
\ion{K}{i} (4047~\AA)& 0.6 $\pm$ 0.2		& 1.0          			\\
C$_2$ (2,0):	&				&				\\
\,R(0) 	        & 0.9$\pm$0.1   	 	& 1.5$\pm$0.1               	\\
\,R(2) 	        & 2.1$\pm$0.1   	 	& 2.2$\pm$0.1               	\\
\,R(4) 	        & 1.8$\pm$0.1   	 	& -		            	\\
\,R(6) 	        & 1.6$\pm$0.1   	 	& -		            	\\
\,R(8) 	        & 1.1$\pm$0.1   	 	& -		            	\\
\,P(4) 	        & 0.9$\pm$0.1   	 	& 0.55                      	\\
\,P(6) 	        & 0.9$\pm$0.1   	 	& -		            	\\
\,P(8) 	        & 0.9$\pm$0.1   	 	& 0.95$\pm$0.15             	\\
\,P(10)	        & 0.7$\pm$0.1   	 	& -		            	\\
\,Q(2) 	        & 2.3$\pm$0.1   	 	& 3.6$\pm$0.1               	\\
\,Q(4) 	        & 2.8$\pm$0.1   	 	& 2.2$\pm$0.1               	\\
\,Q(6) 	        & 2.8$\pm$0.1   	 	& 2.4$\pm$0.1               	\\
\,Q(8) 	        & 1.7$\pm$0.1   	 	& 1.3                       	\\
\,Q(10)	        & 1.2$\pm$0.1   	 	& 0.97$\pm$0.1              	\\
\hline
\label{tb:is-lines}
\end{tabular}
\tablefoot{
\tablefoottext{a}{Non-detections are indicated by ``-'', and indicate either the line is too weak or
contaminated by stellar lines, instrumental artifacts too detect.}
}
\end{table}

\begin{table*}[ht!]
\caption{Equivalent widths (in m\AA) for DIBs in Table~\ref{tb:poleff}}
\label{tb:DIBew}\centering
\resizebox{16cm}{!}{
\begin{tabular}{lll|lll|lll}\hline\hline
   	     \multicolumn{9}{c}{Equivalent widths} \\ \hline
DIB	   &  HD\,197770  &  HD\,194279 &  DIB       &  HD\,197770 &  HD\,194279   &  DIB      &  HD\,197770 &  HD\,194279 \\ \hline
4428.19	   & -	          & -	        &  4979.61   & 4.5$\pm$0.5 & 8$\pm$0.5     &  7045.89  &   2$\pm$1   &	9$\pm$1    \\
5705.08	   & 27$\pm$1 	  & 95$\pm$3	&  5844.92   & 2.5$\pm$0.5 & 6$\pm$0.5     &  7061.05  &   4$\pm$1   & 14$\pm$1	   \\
5780.48	   & 141$\pm$2    & 472$\pm$2	&  5849.81   & 27$\pm$1    & 66$\pm$4	   &  7062.68  &   2$\pm$1   & 12$\pm$1	   \\
5797.06	   & 82$\pm$1     & 149$\pm$3  	&  5973.81   & 3.5$\pm$0.5 & 3$\pm$0.5     &  7069.55  &   8$\pm$1   & 31$\pm$2	   \\
6195.98    & 27.1$\pm$0.5 & 55$\pm$2	&  5975.75   & 6$\pm$0.5   & -  	   &  7077.86  &   -	     &  7$\pm$1	   \\
6203.05	   & 64$\pm$2     & 222$\pm$2 	&  6027.68   & -	   & 18$\pm$1	   &  7119.71  &   22$\pm$1  & 34$\pm$1	   \\
6269.85	   & 40$\pm$2     & 128$\pm$3	&  6065.28   & 5$\pm$0.5   & 9.5$\pm$0.5   &  7375.87  & (1$\pm$1)   &  8$\pm$1	   \\
6283.84	   & -            & -		&  6089.85   &12.5$\pm$0.5 & 18$\pm$0.5    &  7385.89  &   2$\pm$1   &  7$\pm$1	   \\
6379.32	   & 74$\pm$2     & 192$\pm$2 	&  6113.18   & 17$\pm$0.5  & 18$\pm$0.5    &  7405.71  &  2.5$\pm$1  &  4$\pm$1	   \\
6613.62	   & 125$\pm$1 	  & 208$\pm$2  	&  6139.98   & 7.5$\pm$0.5 &  9$\pm$0.5    &  7559.41  &   2$\pm$1   &  9$\pm$1	   \\
6660.71	   & 28$\pm$1	  &  31$\pm$2	&  6234.01   & 9$\pm$0.5   &  16$\pm$0.5   &  7832.89  & 8.5$\pm$1   & 22$\pm$1    \\
6993.13	   & 33$\pm$1	  &  95$\pm$2	&  6244.46   & 9$\pm$1     & -   	   &  7862.43  & (1.5$\pm$1) & (4$\pm$1)   \\
7224.03	   & 51$\pm$2	  & 246$\pm$3 	&  6308.80   &  -	   & (31$\pm$3)    &  8026.25  & (2$\pm$1)   & 32$\pm$1	   \\
8620.41	   & 28$\pm$1	  & 247$\pm$3	&  6317.86   &  -	   &(83$\pm$5)     &           &	     &  	   \\
9577.00    & -     	  &($\sim$120)	&  6425.66   & 10$\pm$1    & 15$\pm$1	   &           &	     &  	   \\
           &		  &		&  6456.01   & 17$\pm$1    & 29$\pm$2	   &           &	     &  	   \\
\hline
\end{tabular}									      
}
\end{table*}

\begin{table*}[ht!]
\caption{Central depths $cd$ (with respect to the normalized continuum) for DIBs in Table~\ref{tb:poleff}.} 
\label{tb:DIBcd}\centering
\resizebox{16cm}{!}{
\begin{tabular}{lll|lll|lll}\hline\hline
   	     \multicolumn{9}{c}{Central depths} \\ \hline
DIB	   &  HD\,197770  &  HD\,194279 &  DIB       &  HD\,197770 &  HD\,194279   &  DIB      &  HD\,197770 &  HD\,194279 \\ \hline
4428.19	   &  $\sim$0.03  & $\sim$0.08	&  4979.61   &  0.005	   & 0.010	   &  7045.89  &    0.005    &  0.013	   \\
5705.08	   &  0.011	  &  0.036      &  5844.92   &  0.006	   & 0.013	   &  7061.05  &    0.013    &  0.028	   \\
5780.48	   &  0.071	  &  0.210      &  5849.81   &  0.034	   & 0.072 	   &  7062.68  &    0.005    &  0.021	   \\
5797.06	   &  0.101	  &  0.168      &  5973.81   &  0.007	   & 0.008	   &  7069.55  &    0.012    &  0.020	   \\
6195.98    &  0.069	  &  0.090      &  5975.75   &  0.019	   & -  	   &  7077.86  &    -	     &  0.012	   \\
6203.05	   &  0.033	  &  0.085      &  6027.68   &  -	   & 0.011 	   &  7119.71  &    0.0145   &  0.025	   \\
6269.85	   &  0.033	  &  0.078      &  6065.28   &  0.011	   & 0.013	   &  7375.87  &   (0.0025)  &  0.013	   \\
6283.84	   &  0.030	  &  0.23       &  6089.85   &  0.026	   & 0.028	   &  7385.89  &    0.005    &  0.012	   \\
6379.32	   &  0.121	  &  0.183      &  6113.18   &  0.025	   & 0.022	   &  7405.71  &    0.004    &  0.008 	   \\
6613.62	   &  0.135	  &  0.179      &  6139.98   &  0.014	   & 0.014	   &  7559.41  &    0.004    &  0.014	   \\
6660.71	   &  0.049	  &  0.045      &  6234.01   &  0.016	   & 0.022	   &  7832.89  &    0.014    &  0.025	   \\
6993.13	   &  0.042	  &  0.101      &  6244.46   &  0.009	   & 0.009	   &  7862.43  &    0.005    &  0.006	   \\
7224.03	   &  0.049	  &  0.199	&  6308.80   &  -	   & 0.013	   &  8026.25  &    0.007    &  0.04	   \\
8620.41	   &  0.007	  &  0.057	&  6317.86   &  -	   & 0.023	   &           &	     &  	   \\
9577.00    &  -		  & $\sim$0.03	&  6425.66   &  0.017	   & 0.020	   &           &	     &  	   \\
           &		  &		&  6456.01   &  0.008	   & 0.025 	   &           &	     &  	   \\
\hline
\end{tabular}									      
}
\end{table*}

\end{appendix}

\newpage
~
\newpage

\begin{appendix}
\section{Stokes $I$, $U$, $Q$, and $V$ spectra for the selected DIBs}

\end{appendix}

\end{document}